\documentclass[prl,twocolumn,showpacs,superscriptaddress,floatfix]{revtex4}
\usepackage{graphicx}

\def\nbZ{{
\mathchoice 
{\hbox{$\sf\textstyle Z\kern-0.4em Z$}}
{\hbox{$\sf\scriptstyle Z\kern-0.3em Z$}} 
{\hbox{$\sf\scriptscriptstyle Z\kern-0.2em Z$}}
}}

\begin{document}

\title{Entanglement in a first order quantum phase transition}
\author{Julien Vidal}
\email{vidal@gps.jussieu.fr}
\author{R\'emy Mosseri}
\email{mosseri@gps.jussieu.fr}
\affiliation{Groupe de Physique des Solides, CNRS UMR 7588,Campus Boucicaut, 140 rue de Lourmel, 75015  Paris, France}
\author{Jorge Dukelsky}
\email{dukelsky@iem.cfmac.csic.es}
\affiliation{Instituto de Estructura de la Materia, CSIC, Madrid, Spain}

\begin{abstract}

The phase diagram of spins 1/2 embedded in a magnetic field mutually interacting antiferromagnetically is determined. Contrary to the
ferromagnetic case where a second order quantum phase transition occurs, a first order transition is obtained at zero field. The
spectrum is computed for a large number of spins and allows one to study the ground state entanglement properties which
displays a jump of its concurrence at the critical point.
\end{abstract}

\maketitle

%
%
%
%
%
%
%
%
%
%

One of the most fascinating feature of the quantum world is certainly the entanglement which has no classical counterpart. Celebrated by the 
pioneering work of Schr\"{o}dinger \cite{Schrodinger}, and Einstein, Podolsky and Rosen \cite{EPR} about the nonlocality,
ubiquitous in the field of quantum information \cite{Preskill,Eckert_book,Nielsen_book}
entanglement properties of quantum systems have recently attracted much attention in the context of
phase transitions.  In the various models studied such as spin chains in a transverse magnetic field
\cite{Osterloh,Osborne,Latorre1,Latorre2,Gu_XXZ,Korepin}, spin ladders \cite{Bose}, spin simplex \cite{Vidal_EntanglementF,Emary}, Hubbard
model
\cite{Gu_Hubbard} the ground state entanglement has been shown to be strongly modified at the critical point raising the question of the
universality of these behaviours.

In this Letter, we consider a system where $N$ spins $1/2$ embedded in a magnetic field $h$ that mutually interact. We focus here on
the antiferromagnetic case, the ferromagnetic one being discussed in Ref. \cite{Vidal_EntanglementF}. The symmetries of the Hamiltonian
allow us to considerably simplify its diagonalization and to determine the phase diagram in the thermodynamical limit. A first order
quantum phase transition is found at zero field whereas in the ferromagnetic case, a second order transition occurs at nonvanishing field.
Next, we study the entanglement properties of the ground state by computing its concurrence which, roughly speaking, measures the two-spin
quantum correlations  \cite{Wootters98}. This concurrence which is nontrivial for $h>0$, is shown to be discontinuous at the transition point
where it switches to zero.

%
%
%
%
%
%
%
%
%
%

We consider the following Hamiltonian which generalizes the model introduced in Ref. \cite{Lipkin}:
%
%
\begin{eqnarray}
H&=&-\frac{\lambda}{N}\sum_{i<j} 
\left( \sigma_{x}^{i}\sigma_{x}^{j} +\gamma\sigma_{y}^{i}\sigma_{y}^{j} \right) 
 -h \sum_{i}\sigma_{z}^{i} \\
&=&-\frac{2 \lambda}{N} \left(S_x^2 +\gamma S_y^2 \right) -2 h S_z + {\lambda \over 2} (1+\gamma),
\end{eqnarray}
%
%
where the $\sigma_{\alpha}$'s are the Pauli matrices and  $S_{\alpha}  =\sum_{i} \sigma_{\alpha}^{i}/2$. The prefactor $1/N$
is necessary to get a finite free energy per spin in the  thermodynamical limit.  
Without loss of generality, we will set $h\geq 0$ in the following.
The Hamiltonian $H$ preserves the magnitude of the total spin and does not couple states having a different parity of
the number of spin pointing in the magnetic field direction (spin-flip symmetry), namely:
%
%
\begin{eqnarray}
[H,{\bf S}^2]&=&0,\\
\left[H,\prod_i \sigma_z^i \right] &=&0,
\label{phaseflip}
\end{eqnarray}
%
%
for all anisotropy parameter $\gamma$. In addition, it is straightforward to show that the full spectrum of $H$ is odd under the
transformation  $\lambda \rightarrow -\lambda$ and even under $h \rightarrow -h$ .  Furthermore, since $H$ writes in terms of the
total spin operators, the degeneracy of each eigenvalues belonging to a spin $S$ sector is at least equal to the number of spin $S$
representations which is simply given by~:
%
%
\begin{eqnarray}
D_S&=&\left(
\begin{array}{c}
N
\\
N/2-S
\end{array}
\right)-
\left(
\begin{array}{c}
N
\\
N/2-S-1
\end{array}
\right)
\end{eqnarray}
%
%
for all $N$. This implies that the {\em full} spectrum is obtained by diagonalizing only one representation of each spin sector $S$
which allows us to deal with a large number of spins.  Denoting by $\{|S,M\rangle \}$ an eigenbasis of ${\bf
S}^2$ and $S_z$, the matrix elements of $H$ reads~:
 %
%
\begin{eqnarray}
\langle S',M' |H|S,M\rangle&=&\delta_{S ,S'} \Bigg\{ \left[ -{\lambda \over N}(1+\gamma) \left( S(S+1 \right) -M^2 \right. \nonumber\\
&& -N/2 \left. \right)-2h M  \bigg] \delta_{M',M}- {\lambda (1-\gamma) \over 2 N} \nonumber\\
&& 
\left( a_{M-1}^{S^-} a_{M}^{S^-} \delta_{M',M-2} \right. + \nonumber\\
&& 
\left. a_{M+1}^{S^+} a_{M}^{S^+} \delta_{M',M+2}\right) \Bigg\}
\end{eqnarray}
%
%
where $a_{M}^{S^\pm}=\sqrt{S(S+1)-M(M \pm 1)}$
These expressions which are valid for any value of the parameters $(\lambda, h, \gamma)$ and for any $N$ generalize to any spin sector
those given in Ref. \cite{Vidal_EntanglementF} for the Dicke subspace ($S=N/2$).  In the isotropic case  $\gamma=1$, one further has
$[H,S_z]=0$ so that $H$ is diagonal in the basis  $\{ |S,M \rangle \}$.

The antiferromagnetic nature of the coupling between spins considered here ($\lambda<0$) completely modifies the phase diagram of $H$ as
compared to the ferromagnetic case. A simple mean-field approach analogous to the one presented in Ref. \cite{Botetprl,Botetprb} can
be performed and predicts a first order phase transition in the zero field limit for any positive $\gamma$.  
The magnetization (per spin) in the $z$ direction of the ground state is given for all $\gamma \geq 0$ by:
%
%
\begin{equation}
{1\over N} \langle S_z \rangle = {1\over 2} {\rm sgn}(h)
\end{equation}
%
%
where sgn denotes the sign function which vanishes for $h=0$. In the thermodynamical limit, the ground state is thus the fully polarized
state for $h\neq0$. Contrary to the ferromagnetic case, the ground state for a finite arbitrary $N$ does not, {\it a priori}, lie in
the symmetric representation of the permutation group spanned by the Dicke states \cite{Vidal_EntanglementF}. Indeed, 
in the zero field limit ($h\rightarrow 0$), the ground state shall belong to the lowest spin sector that
minimizes the interaction term. The main issue is thus to determine the values of $h$ for which level crossings appears.

%
%
\begin{figure}[ht]
\includegraphics[width=96mm]{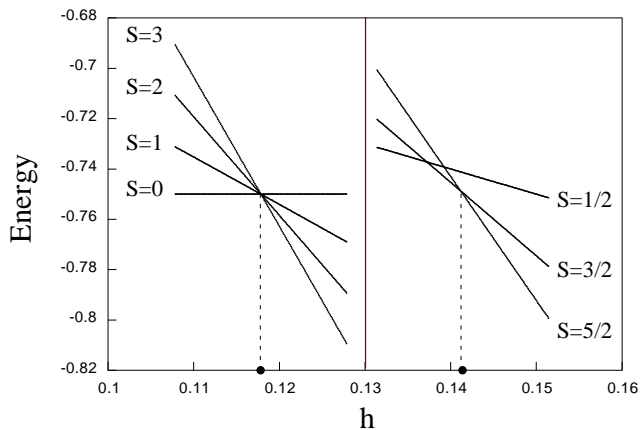}
\vspace{-48mm}
\caption{Ground state energy of each spin sector as a function of the magnetic field for $\gamma=1/2$ ($\lambda=-1$). In the even case $N=6$
(left) all level degenerate at $h_{\rm susy}$ ($\bullet$) whereas in the odd one $N=5$ (right) a cascade is observed from the
$S=N/2$ to $S=1/2$ sector.}
\label{Ground}
\end{figure}
%
%

For illustration, we have displayed in Fig. \ref{Ground} the ground state energy of the different spin sector for a small number of spins. 
Two different scenarios arise according to the parity of $N$.
In the even $N$ case an additional symmetry  allows us to give a complete description of the ground state properties. Indeed, as recently
pointed out by R.~G. Unanyan and M. Fleischhauer \cite{Unanyan}, when 
%
%
\begin{equation}
|h_{\rm susy}|={|\lambda| \sqrt{\gamma} \over N} ,
\label{condition}
\end{equation}
%
%
the Hamiltonian is supersymmetric \cite{Witten81}. When the SUSY condition (\ref{condition}) is fulfilled for an
antiferromagnetic interaction ($\lambda<0$), the authors of Ref. \cite{Unanyan} claim that the ground state is nondegenerate and given
by~: 
%
%
\begin{equation}
|\psi_0 \rangle=A e^{-\eta S_z}|N/2,0_y\rangle,
\label{groundSUSY}
\end{equation}
%
%
where $A$ is a normalization constant and where $|N/2,0_y\rangle$ denotes the eigenstate of ${\bf S}^2$ and $S_y$ with eigenvalues
$(N/2)(N/2+1)$ and
$0$ respectively. The parameter $\eta$ is determined by the relation \cite{Error}~: $\tanh \eta=\lambda/N$. 
Even though $|\psi_0 \rangle$ is a ground state of $H$ with total spin $S=N/2$, the spectrum at the SUSY point is however {\em highly
degenerate}.  Indeed, at the SUSY point $h_{\rm susy}$, all the lowest eigenvalues of each spin
representation are equal ($E_0=\lambda (1+\gamma)/2)$ and the degeneracy of the ground state is thus given by~:
%
%
\begin{equation}
d_g(\lambda=\lambda_{\rm susy})=\sum_{S=0}^{N/2} D_S=
\left(
\begin{array}{c}
N
\\
N/2
\end{array}
\right).
\end{equation}
%
%
This collapse of the spectrum can be easily analyzed in the isotropic case $\gamma=1$ for which any state $|S,M \rangle$ is eigenstate
with eigenvalue~:
%
%
\begin{equation}
E(S,M)=  -{2 \lambda \over N} \left( S(S+1) -M^2 \right) +\lambda -2h M.
\end{equation}
%
%
For $\lambda<0$ and $h>0$, $E(S,M)$ is minimum for $M=S$ and at SUSY point $E(S,S)=\lambda$ for any $S$.

The existence of this supersymmetric point enables us to locate the ground state for any value of the parameters.
Indeed, $H$ describes a competition between the magnetic field $h$ which aims to align the spin in the field direction, and the interaction
terms which favors antiferromagnetic configurations. Thus, since at the SUSY point there
exists a ground state lying in the maximum spin sector $S=N/2$, the ground state for $h>h_{\rm susy}$ also
lies in this sector and $d_g(h>h_{\rm susy})=D_{N/2}=1$. 
Similarly, since singlet states $|0,0\rangle$ are also ground states at the SUSY point, they remain ground state for $h<h_{\rm susy}$ and 
$d_g(h<h_{\rm susy})=D_{0}$. 
In the thermodynamical limit ($N \rightarrow \infty$), $h_{\rm susy}$ goes to zero and the level crossing
between $S=N/2$ and $S=0$ ground states thus occurs at zero field. Nevertheless, for $h=0$, it is clear that the ground states has a zero
total spin for all $\gamma\geq 0$.

For odd $N$ , the situation is more complex since the ground states of each spin sector do not degenerate as in the even $N$ case except
in the thermodynamical limit. In fact, as previously, the ground state belongs to the $S=N/2$ sector for $h>h_{\rm susy}$, and
then switches to the other spin sector with decreasing $S$  when $h$ is lowered below $h_{\rm susy}$. 
Naturally, for $h=0$, the ground state lies in the minimum spin sector and is given by all states $|1/2,1/2 \rangle$. 
Its degeneracy $D_S$ thus strongly depends on $h$. Nevertheless, in the thermodynamical limit, the region in which these level crossings
occurs shrinks and converged to the zero field point so that the parity of $N$ becomes irrelevant for the macroscopic physical  quantities. 

%
%
%
%
%
%
%
%
%
%

To analyze the entanglement properties of the ground state, we focus on the concurrence $C$
which has been introduced by Wooters \cite{Wootters98} to measure the two-spin entanglement. This quantity 
is obtained from the density matrix describing the state to be characterized.  Here, we concentrate on the ground
state of $H$ that can, as explained above, be degenerate. Thus, we must consider the thermal density
matrix (at zero temperature) defined by~:
%
%
\begin{equation}
\rho_{\rm th.}={1 \over d_g} \sum_{i=1}^{d_g} |\psi_i\rangle \langle \psi_i|,
\end{equation} 
%
%
where $d_g$ is the degeneracy of the ground state and where $\{|\psi_i\rangle,i=1,\dots,d_g\}$ constitutes an {\em orthogonal} basis of the
$d_g$-dimensional lowest energy subspace ${\cal E}_0$. Indeed, if we would consider the projector onto a
specific state belonging to ${\cal E}_0$, the entanglement properties would strongly depend on this choice.
Then, let
$\rho$ be the reduced density matrix obtained by tracing out
$\rho_{\rm th.}$ over $(N-2)$ spins. Of course, in our system, the choice of the two spins kept is irrelevant because of the permutation
symmetry. Next, we introduce the spin-flipped density matrix  
$\tilde \rho=\sigma_{y}\otimes\sigma_{y}\:\rho^{\ast}\:\sigma_{y}\otimes\sigma_{y}$ where 
$\rho^{\ast}$ is the complex conjugate of $\rho$. The concurrence $C$ is then defined by:
%
%
\begin{equation}
C=\max\left\{ 0,\mu_{1}-\mu_{2}-\mu_{3}-\mu_{4}\right\},
\end{equation}
%
%
where the $\mu_{j}$ are the square roots of the four real eigenvalues of $\rho \tilde{\rho}$, classified in decreasing order. 
This concurrence vanishes for an unentangled two-body state whereas $C=1$ for a maximally entangled one. As explained in Ref.
\cite{Vidal_EntanglementF}, it is further important to deal with a rescaled concurrence  $C_R=(N-1) C$ to take into account 
the coordination number of each spin.
For a large number of spins, the difficulty comes from the trace step which requires operation in the full
Hilbert space which is $2^N$-dimensional. 

For $h>h_{\rm susy}$, this can be achieved since the ground state lies in the sector $S=N/2$. Indeed, (i) the $S=N/2$ subspace is
nondegenerate so that the thermal density matrix of the unique ground state \cite{degeneracy} $|\psi\rangle=\sum_M \alpha_M |N/2,M\rangle$
simply writes $\rho_{\rm th.}=|\psi\rangle \langle\psi|$, and  (ii) the symmetry of the Dicke states $|N/2,M\rangle$ \cite{Dicke}
allows us to write down in a simple form the reduced density matrix $\rho$ in the standard basis
$\{|\uparrow \uparrow \rangle, |\uparrow \downarrow \rangle,|\downarrow \uparrow \rangle,|\downarrow \downarrow \rangle 
\}$. One has \cite{Stockton}:
%
%
\begin{eqnarray}
\rho_{11}&=& \sum_M |\alpha_M|^2 {(N+2M)(N+2M-2) \over 4N(N-1)} \\
\rho_{22}&=& \sum_M |\alpha_M|^2 {(N-2M)(N+2M) \over 4N(N-1)} \\
\rho_{44}&=& \sum_M |\alpha_M|^2 {(N-2M)(N-2M-2) \over 4N(N-1)} \\
\rho_{14}&=& \sum_M \alpha_M \alpha_{M+2}^* \sqrt{(N+2M)(N+2M-1)} \times \nonumber\\
&&{\sqrt{(N-2M+1)(N-2M+2)} \over 4N(N-1)}.
\end{eqnarray} 
%
%
Furthermore, one has $\rho_{23}=\rho_{33}=\rho_{22}$, and $\rho^*=\rho$. The other matrix elements can also be 
computed for an arbitrary state belonging to $S=N/2$ subspace but since they vanish for the eigenstates of $H$ because of
(\ref{phaseflip}), we do not give them here. Note that for the  Dicke states, one recovers the expressions given in Refs.
\cite{Wang-Pairwise,Stockton}.

For $h<h_{\rm susy}$, the ground state lies in the minimum spin sector and is highly degenerate. The thermal density matrix is then simply 
the projector onto the subspace corresponding to $S=0$ for even $N$ and $S=1/2$ for odd $N$. 
The concurrence computed from such a  density matrix is shown to vanish for $N>3$ \cite{prep}.  

\begin{figure}[ht]
\includegraphics[width=100mm]{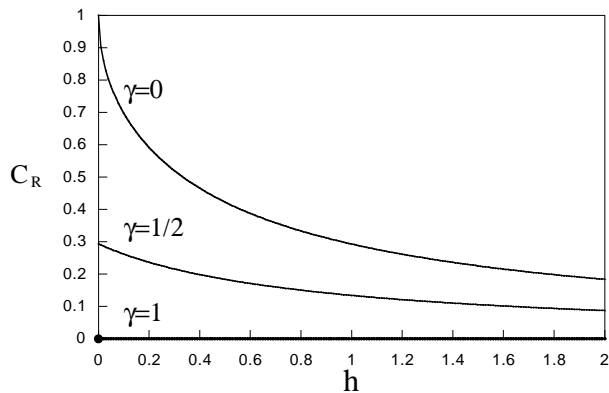}
\vspace{-15mm}
\caption{Rescaled concurrence of the ground state as a function of the magnetic field for various anisotropy parameter $\gamma$ and
for $N=10^3$ spins $(\lambda=-1)$. Note that for any $\gamma$, one has $C_R=0$ at zero field.}
\label{Conch}
\end{figure}

Figure \ref{Conch} shows the behaviour of the rescaled concurrence of the ground state as a function of the magnetic field. 
Note that for $\gamma=1$ and $h>h_{\rm susy}$ the ground state is simply the Dicke state $|N/2,N/2\rangle$ for which $C=0$. The same 
result also trivially holds in the large $h$ limit for any $\gamma$.

%
%
%
%

\begin{figure}[ht]
\includegraphics[width=100mm]{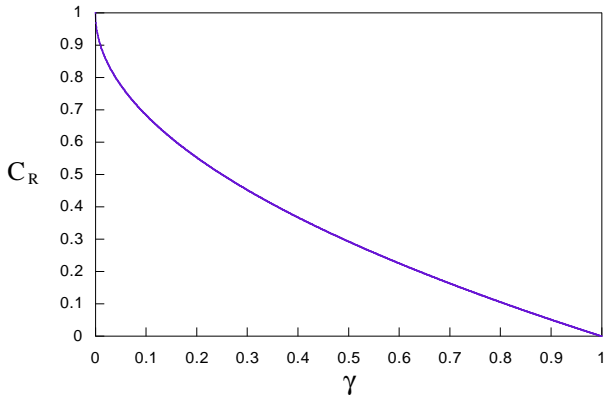}
\vspace{-15mm}
\caption{Rescaled concurrence of the ground state $|\psi_0\rangle$ at the SUSY point as a function of the anisotropy
parameter
$\gamma$ for $N=10^4$ spins.}
\label{concSUSY}
\end{figure}

For $h=h_{\rm susy}$ the worse situation is reached, at least for even $N$, since there is one ground state in each spin sector. As a
consequence, the thermal density matrix is a sum of projector onto states of very different nature so that the trace operation is a rather
difficult task for large $N$.  
Here, we have chosen to focus on $|\psi_0 \rangle$ whose analytic expression (\ref{groundSUSY}) allows one to
compute its concurrence for any even $N$. The coefficients $\alpha_M$ entering in its decomposition onto the Dicke states are simply given
by~:
%
%
\begin{equation}
\alpha_{N/2-2j}=e^{-\eta (N/2-2j)}
\left(
\begin{array}{c}
N/2
\\
j
\end{array}
\right)
\Bigg/
\left(
\begin{array}{c}
N
\\
2j
\end{array}
\right)^{1/2}
\end{equation}
%
%
where we have set $M=N/2-2j$ because of the symmetry $\prod_i \sigma_z^i |\psi_0 \rangle=|\psi_0 \rangle$. 
Two limiting cases can be easily analyzed~: the $XY$ case ($\gamma=1$) for which $C_R=0$, and the Ising model ($\gamma=0$) for
which
$\eta=0$ and $C_R=1$. 
The rescaled concurrence of $|\psi_0 \rangle$ as a function of $\gamma$ is displayed in Fig. \ref{concSUSY}. 
%
%
%
%

In the large $N$ limit, the behaviour of the rescaled concurrence of $|\psi_0\rangle$ can be computed and is given by~:
%
%
\begin{equation}
C_R(\gamma)=1-\sqrt{\gamma}
\end{equation}
%
%
for $0\leq \gamma \leq 1$. Of course, this information does not enable us to conclude anything about the concurrence computed with the full
thermal density matrix involving all spin sector but it certainly points out a nontrivial behaviour of the true ground state at 
$h=h_{\rm susy}^+$.

Such a discontinuity of the concurrence at the transition point has already been obtained in other frustrated spin models  such as spin
ladders or Heisenberg antiferromagnets in the Kagom\'e lattice \cite{Bose}. 
However, in these systems the ground state entanglement properties are very simple (valence-bond like states) so
that its concurrence is constant in each region of the phase diagram.
If a jump of the ground state concurrence seems reasonable for a system displaying a first order quantum phase transition, it is not obvious
that other measures of the entanglement would have shown discontinuity. More precisely, in the above mentioned example as well as in our
model, the concurrence is found to become trivial for some parameter values. An interesting perspective would be to study other measures of
the entanglement in systems displaying a quantum phase transition such as the $N$-tangle \cite{Wong} or the Minkovskian-square norm of the
Stokes tensor \cite{Jaeger1_pra} which investigate the $N-$spin entanglement. Indeed, the trace operation performed in the concurrence
calculation undoubtedly kills some correlations between spins that could be captured by other types of measurement. \\

\acknowledgments
We are indebted to B. Dou\c{c}ot and D. Mouhanna for fruitful and valuable discussions.


\begin{thebibliography}{10}
\bibitem{Error} Note that a sign error has been inserted in Ref. \cite{Unanyan}.
\bibitem{degeneracy} Contrary to the ferromagnetic case where it can be doubly degenerate \cite{Vidal_EntanglementF}, the
ground state in the $S=N/2$ sector is always nondegenerate.
\end{thebibliography}




\end{document}